# In situ annealing effects on the iron-based ladder material BaFe$_2$S$_3$: A route to improve the crystal quality

Xuan Zhang[1,2], Hui Zhang[2,3], Yonghui Ma[2,4], Lingling Wang[2], Jianan Chu[2,4], Tao Hu[2,3], Gang Mu[2,3*], Yuming Lu[1], Chuanbing Cai[1], Fuqiang Huang[3,5], and Xiaoming Xie[2,3]

**ABSTRACT** We have grown single crystals of the iron-based ladder material BaFe$_2$S$_3$, which is superconductive under high pressure, adopting different conditions. By comparing the behaviors of these samples, it is found that the in situ annealing process can affect the crystal structure and the electrical transport, enhance the antiferromagnetic transition temperature, and reduce the extrinsic ferromagnetic component of the system. An in-depth analysis indicates that the crystal quality is improved by the in situ annealing in terms of reducing both the Fe deficiency and the Fe impurity in the samples. The improvement of the sample quality will facilitate the investigations on the intrinsic properties of this material.

## INTRODUCTION

The iron-based superconductors, regardless detailed crystal structure, all share the same conducting Fe$_2$X$_2$ (X = As, P, Se and S) layers as the crucial structural unit [1-8]. In the Fe$_2$X$_2$ layers, edge-shared FeX$_4$ tetrahedrons form the square lattice in the quasi-two-dimensional (quasi-2D) planes. One question is what will happen in the quasi-one-dimensional (quasi-1D) materials with the same coordination as the quasi-2D cases. Theoretically proposals on the magnetic states and possible superconductivity have been reported in the recent five years [9-11]. Recently, a quasi-1D compound BaFe$_2$S$_3$, with the same FeX$_4$ tetrahedrons extending along only one direction, were studied and found to be superconductive under the pressure around 11 GPa [12, 13]. The maximum superconducting transition temperature T$_c$ can be as high as 24 K. Moreover, superconductivity was found to locate adjacent to the antiferromagnetic (AFM) state in the phase diagram, similar to the unconventional superconducting systems like cuprates, iron-based superconductors, heavy-fermion superconductors, and so on [14]. In addition, this material is known as a Mott insulator with strong electron correlation effect [11-13]. These preliminary experimental facts put the compound BaFe$_2$S$_3$ on the position of a possible new candidate member of the unconventional superconducting family. However, it is found that the occurrence or not and the quality of the superconducting transition are closely associated with the sample qualities. Consequently, an in-depth investigation on the single crystal growth and quality improvement is crucial for studying the intrinsic properties of this interesting system. At present, such investigations are still lacking although tuning of the electrical properties by doping has been carried out [15].

In this paper, by comparing the behaviors of the samples grown with two different conditions, we discriminated the key factors that affect the sample quality and seek out a route to improve it. It is found that both the Fe deficiency and the Fe impurity can be reduced by an in situ annealing process, based on which the intrinsic resistivity behavior is approaching.

## EXPERIMENTAL SECTION

BaFe$_2$S$_3$ single crystals were synthesized by a melting method. Two different post processes, namely water quenching and in situ annealing, were applied. Commercially BaS (powder, Alfa Aesar, 99.7%), Fe (powder, Alfa Aesar, 99.998%) and S (powder, Alfa Aesar, 99.5%) were mixed with mole ratio of 1:2:2. The powder mixture was ground, pressed into a pellet shape, and placed into a carbon crucible, which was then sealed in a quartz tube. The tube was heated up to 300 ℃ and held for 10 hours to prevent S powder from volatilizing so quickly. Next, the tube was heated up to 1100 ℃ at the rate of about 70 ℃/h, kept 24 h, and then slowly cooled to 750 ℃ over 24 h. For the water quenching condition, the quartz tube was filled with high purified helium of about 0.3 atm, and quenched by quickly dropping the tube into water after the cooling to 750 ℃. While for the in situ annealing condition, the quartz tube was evacuated to vacuum, and kept at 500 ℃ for 5 days at the last stage.

The structure of the obtained samples was checked using a DX-2700 type powder x-ray diffractometer (PXRD). The crystals were crushed be the PXRD measurements. The magnetic susceptibility measurements were carried out on the magnetic property measurement system (Quantum Design, MPMS 3). The electrical resistance was measured using a four-probe technique on the physical property measurement system (Quantum Design, PPMS).

## RESULTS

We examined the structure and purity of our samples by the powder x-ray diffraction measurements at room temperature. In Fig. 1b, we show the PXRD data of the crushed crystals for both the quenched and annealed samples. All the diffraction peaks can be indexed to the orthorhombic ladder structure with the space group Cmcm as shown in Fig. 1a. No impurity can be detected from PXRD data for the both samples. We note that the grain orientation of the crushed crystals is still highly anisotropic since the relative intensity of the diffraction peaks is different from the polycrystalline samples [16]. Nevertheless, the information of the lattice constants can still be derived from the position of the diffraction peaks. As shown in Fig. 1c, the peak of the annealed sample move to the left slightly as compared with the quenched sample, indicating the expansion of crystal lattice induced by the annealing process. Quantitatively, the lattice constants for the quenched and annealed samples are a = 8.7705 Å, b = 11.2180 Å, c = 5.2822 Å, and a = 8.7794 Å, b = 11.2252 Å, c = 5.2853 Å respectively. The lattice expands along three directions and the unit cell volume is enhanced by about 0.2% for the annealed sample.

Temperature dependence of magnetic susceptibility $\chi$ is shown as the solid symbols in Fig. 2a. The data were measured under the field 5 T along the c axis (the legs of the ladder) of the samples. Clear transition can be seen at around 110 K, which is consistent with the previous report and has been attributed to the stripe-type antiferromagnetic (AFM) transition [12]. The roughly linear increase of $\chi$ with warming in the high temperature side is similar to that observed in the parent compounds of cuprates and iron-based superconductors [17, 18], and may be a feature of short-range AFM correlations [19]. Here we concentrate on the comparison between the two samples grown under two different conditions. We found that the $\chi$ value of the quenched sample is clearly larger than that of annealed sample. We emphasize that such a tendency is reproducible because several different samples grown under the same condition were measured and consistent results were obtained. Moreover, the Neel temperature $T_N$ is slightly different for the two samples. As shown in Fig. 2b, the normalized data near the AFM transition indicate that $T_N$ is enhanced for 7 K by the annealing process compared with the quenched sample. When checking the field dependence of the moment as shown Figs. 2c and d, the nonlinear behavior with hysteresis, which is the characteristic of the ferromagnetic (FM) ordering, can be observed in the low field region, while an unsaturated linear increase of the moment with field is revealed in the high field region. Such a behavior emerges in the temperature region from 2 K to 300 K, indicating a rather high Curie temperature above 300 K for the FM ordering. An intuitional and reasonable understanding for this observation is that a FM component coexists with the AFM phase (in the low temperature) and the short-range AFM region (in the high temperature). A linear fit to the data of the high-field linear region, as revealed by the dashed lines in Fig. 2c and d, can extract the contributions excluding the FM component. The extracted data are displayed by open symbols in Fig. 2a. The magnitudes of the quenched and annealed samples are very close to each other, suggesting that the intrinsic magnetic property of this material is obtained in this way. The differences in the raw data (see the solid symbols in Fig. 2a) come from the FM component, which seems to be extrinsic.

The properties of electrical transport were studied by measuring the temperature dependence of resistivity. The current is along the legs of the ladder. We found that the normalized resistivity $\rho/\rho_{300K}$ shows a better reproducibility than the raw data when comparing the data from different samples of the same growth condition, from which one can catch the characteristics of the electron scattering. The annealed sample shows a stronger insulating behavior with cooling compared with the quenched one (see Fig. 3a). Being consistent with the previous report, the thermal activation model could not describe our data [13, 15]. Instead, the temperature dependence of resistivity is characterized by the variable-range hopping conduction (VRH), $\rho \sim \exp[(T_B/T)^{1/(1+d)}]$. Here $T_B$ is a constant and d is the dimensionality. In Figs. 3b and c, we plot our data as $\log(\rho/\rho_{300K})$ vs $T^{-1/(1+d)}$ with d = 1 and 2 respectively. Actually, the apparent difference between the 1D and 2D VRH is rather subtle. After a scrutinization on the low temperature data one would find that the annealed sample is more coincident with the 1D VRH, while the 2D VRH is more suitable for the quenched sample.

## DISCUSSION

We have summarized the comparison of the behaviors between the quenched and annealed samples in Table 1.

Based on these experimental results, we are able to achieve a comprehensive understanding. Our results verify the speculation proposed by Takahashi et al. [12] (in the supplementary information) that the presence of Fe deficiencies lowers the $T_N$ values. In addition, we find that the unreacted Fe gives rise to the FM component in the magnetic measurements, which is reduced and fills in the positions of Fe deficiencies in the annealing process. The decrease of Fe deficient enhances both the lattice constants and $T_N$. All these tendencies are reasonable and consistent with the experimental facts (1), (2), and (3) shown in Table 1. In the work of Takahashi et al. [12], an excess of Fe was added to remove the Fe deficiencies and enhance $T_N$. However, such a treatment also increases the amounts of unreacted Fe impurities and the FM component, which can be grasped by examining the decline degree of the magnetic susceptibility from 150 K to $T_N$ in the normalized scale (see Figure S4 of Ref. [12]) since the decline degree will be larger for the sample with fewer FM component as revealed in Fig. 2b. Obviously, the in situ annealing method adopted in our work reduces the Fe impurities and Fe deficiencies of the samples simultaneously, which is a promising approach for improving the crystal quality and may be beneficial to the investigation on the intrinsic properties of the present system.

One example is the transport behaviors shown in Figs. 3b and c. 1D VRH is more natural and reasonable for the present material with quasi-1D crystal structure. The high level of imperfection in the quenched sample may prevent the observation of such an intrinsic property. More experiments tuning the annealing temperature and time of duration are needed and underway to further enhance the sample quality.

## CONCLUSIONS

We have investigated the crystal growth and quality characterization of the iron-based ladder material $BaFe_2S_3$. It is found that the unreacted Fe induced impurity and Fe deficiencies are two main factors reducing the crystal quality, which can be suppressed simultaneously by an in situ annealing process. Our results supply an effective route to improve the crystal quality of the present system. Moreover, in the annealed samples with a higher quality, the electrical transport shows a 1D variable-range hopping feature in the low temperature region.

**Acknowledgements** This work was financially supported by the Youth Innovation Promotion Association of the Chinese Academy of Sciences (No. 2015187), the National Natural Science Foundation of China (No. 11204338) and the "Strategic Priority Research Program (B)" of the Chinese Academy of Sciences (No. XDB04040300).

**Author contributions** Mu G designed the experiments. Zhang X synthesized the samples and conducted the magnetic and electrical transport measurements. Zhang H analyzed the crystal structure. Zhang X and Mu G wrote the paper. All authors contributed to the general discussion.

**Conflict of interest** The authors declare that they have no conflict of interest.

**Supplementary information** Experimental details are available in the online version of the paper.

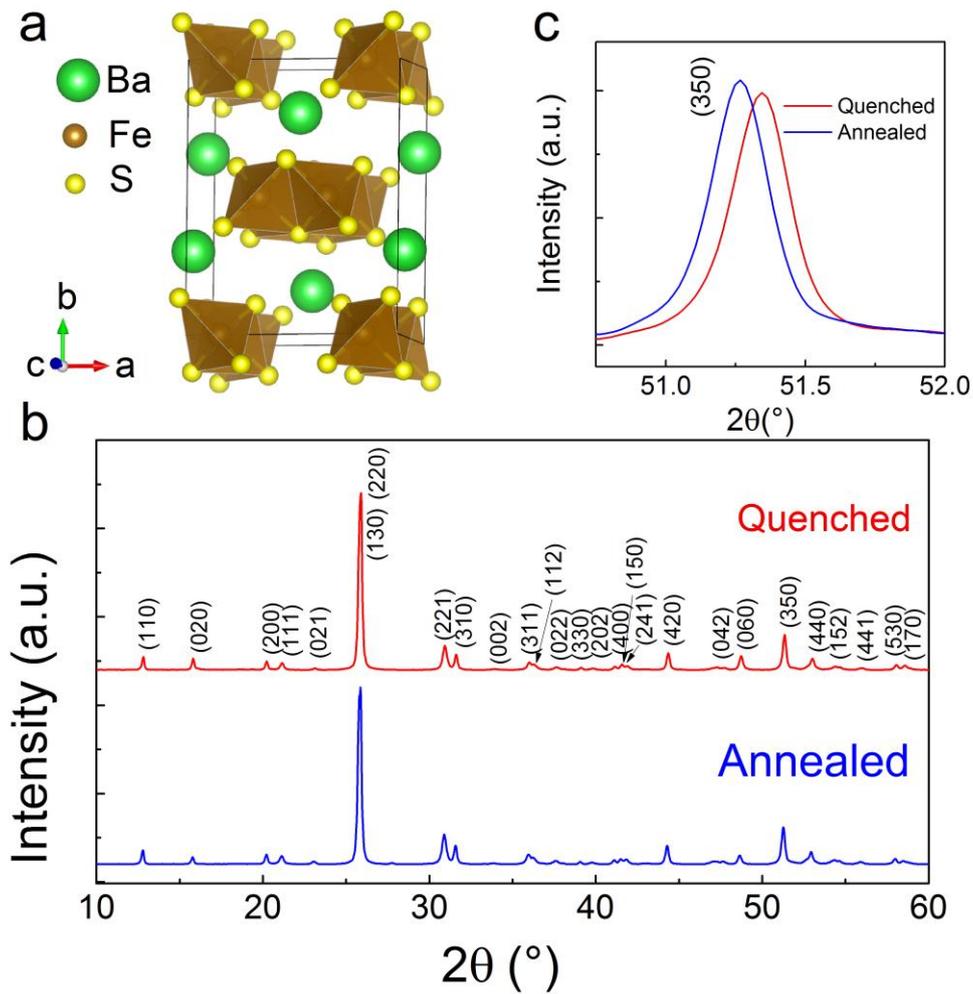

**Figure 1** (a) The schematic of the crystal structure of $BaFe_2S_3$. (b) Powder x-ray diffraction patterns for two samples with the quenching and annealing treatments. The powder samples were obtained by crushing the crystals. (c) An enlarged view of the PXRD patterns near (350) peak.

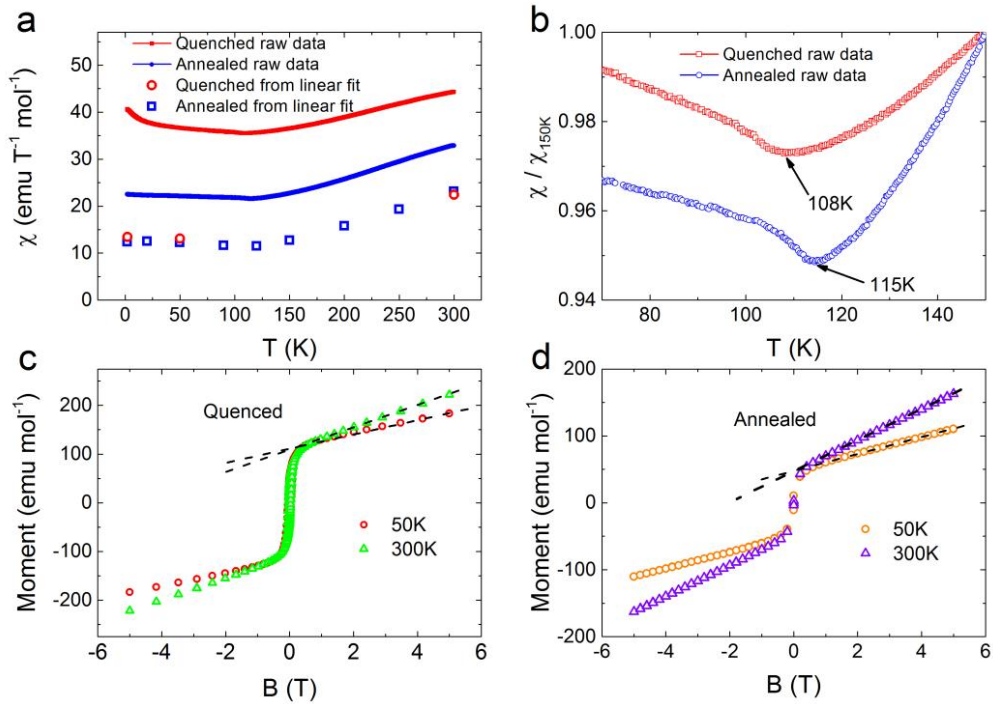

**Figure 2** (a) Temperature dependence of the magnetic susceptibility for two samples with the quenching and annealing process. The solid symbols are raw data, while the open symbols are obtained from the linear fit of the field dependent data shown in (c) and (d). (b) An enlarged view of the normalized magnetic susceptibility data near the AFM transition. (c) – (d) Field dependence of the moment for quenched and annealed samples respectively. The dashed lines are linear fits to data in the high field region.

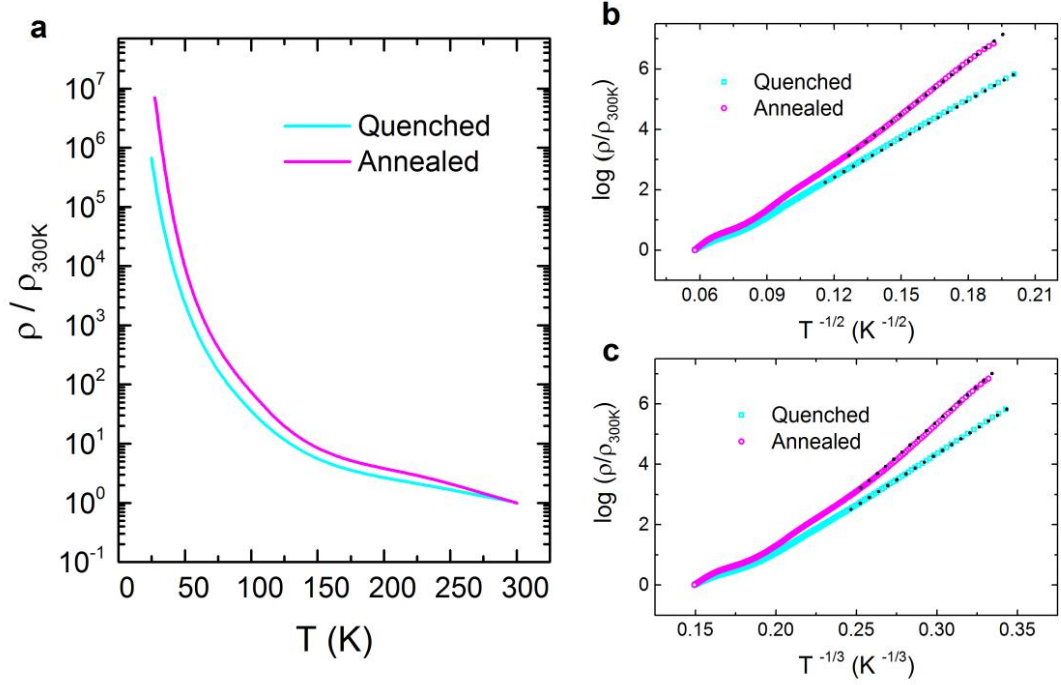

**Figure 3** (a) Temperature dependence of the normalized resistivity for the quenched and annealed samples. (b) – (c) Plot with $\log(\rho/\rho_{300K})$ vs $T^{-1/(1+d)}$ (d = 1 and 2) to check the 1D and 2D VRH models. The dotted lines are the linear fits to the data in the low temperature region.

**Table 1** Comparison of the behaviors between the quenched and annealed samples.

|          | (1)              | (2)              | (3)          | (4)                 |
|----------|------------------|------------------|--------------|---------------------|
| Sample   | Lattice constant | Neel temperature | FM component | Resistivity behavior |
| Quenched | Small            | Low              | High         | 2D VRH              |
| Annealed | Large            | High             | Low          | 1D VRH              |